\documentstyle[11pt,newpasp,epsf]{article}

\begin{document}

\title{Magneto--rotational and Thermal Evolution of Magnetars with Crustal 
Magnetic Fields }
\author{U. Geppert}
\affil{Astrophysikalisches Institut Potsdam, An der Sternwarte 16, 
       14482 Potsdam, GERMANY, e-mail: urme@aip.de}
\author{D. Page}
\affil{Instituto de Astronom\'{\i}a, UNAM, 
       04510 Mexico D.F., MEXICO, e-mail: page@astroscu.unam.mx}
\author{M. Colpi}
\affil{Dipartimento di Fisica, Universit\'a degli Studi di Milano Bicocca,
       Via Emanueli 15, 20133 Milano, Italy, colpi@uni.mi.astro.it} 
\author{T. Zannias}
\affil{Instituto de F\'{\i}sica y Matem\'aticas, Universidad Michoacana SNH, 
       Morelia, Mich. 58040, MEXICO, e-mail: zannias@ginette.ifm.umich.mx}

\section{Introduction}

The interpretation of Soft--Gamma--Repeaters (SGRs) and Anomalous X--Ray Pulsars (AXPs) as Magnetars (Thompson \& Duncan 1996) 
raises again
the issue  of the generation of the ultra--strong magnetic fields (MFs) 
in neutron stars (NSs) and the related question of where these fields are 
anchored:
in the core, penetrating the whole star, or confined to the crust.
Recently, Heyl \& Kulkarni (1998) considered the magneto--thermal evolution of 
magnetars
with a core field. Since the assumption of a crustal field is at least not in 
disagreement with the observations of isolated pulsars (Urpin \& Konenkov 
1997)
and of  NSs in binary systems (Urpin, Geppert \& Konenkov 1998, Urpin, 
Konenkov \& Geppert 1998), here we would like to address the question whether 
the
observations of SGRs and AXPs can be interpreted as magnetars having a crustal 
MF.
Given the strength of the MF in magnetars we take into account, in an 
approximate manner, the strongly non--linear Hall effect on its decay.
We intend to provide a contribution to an unified picture of NS MF evolution 
based
on the crustal field hypothesis.

\vspace{-0.2cm}
\section{Model}

We consider three qualitatively different equations of state (EOS), a stiff 
one from
Pandharipande et al. (1976), a medium one from Wiringa et al. (1988)
and a soft one from Pandharipande (1971) which includes the effect of 
hyperons.
We consider for all EOSs the possibility of both slow neutrino emission from 
the modified
Urca processes and fast neutrino emission by the direct Urca processes;
all standard neutrino emission processes in the crust are also properly taken 
into account
(see, e.g., Page, 1998). 
The thermal evolution is essentially affected by the different neutrino 
emissivities,
however, in the case of magnetars Joule heating and the effect of the strongly 
magnetized
envelope modify the thermal history considerably compared to that of standard 
NSs.
The Joule heating rate is determined by $j^2/\sigma_{\rm eff}$, where $j$ is 
the current
density and $\sigma_{\rm eff}$ the electrical conductivity (see below).
Since the calculation is performed under spherical symmetry this value is 
averaged
over spheres of constant radius.
Ultra--intense MFs affect strongly the transport of heat in the upper layers 
of the star,
the envelope, resulting in a higher flux than in the case of weaker fields 
(Heyl \& Hernquist 1998). 
This implies that young magnetars will be naturally hotter than standard young 
NSs
even if their interior followed the same thermal path.
We, moreover, assume here that the envelope is made of iron: an envelope 
containing
light elements would lead to an even higher flux. 
We solve the general relativistic equations of heat conservation and transport 
in the
whole star by using a Henyey type code developed specially for NS cooling 
studies
(see Page 1998 and references therein).
We consider the evolution of a purely poloidal dipolar MF, maintained by 
currents in
the crust of the NS. 
It is governed by
\vspace{-0.2cm}
\begin{equation}
\frac{1}{c}\frac{\partial \vec B}{\partial t} + \vec{\nabla}\wedge 
 \left[ \frac{c}{4\pi \sigma_{\rm eff}}          \vec{\nabla}\wedge  
       \left(e^{\Phi}\vec B  \right)
\right] =0
  \;\;\;\;\;\; {\rm with} \;\;\;\;\;\;
  \sigma_{\rm eff} = \sigma_{\rm eff}(T,\rho,\vec B,...)
\vspace{-0.1cm}
\end{equation}
The Hall effect will not directly decay the MF, but will redistribute 
magnetic
energy into field modes with smaller length scales (the {\em HALL CASCADE}, 
Goldreich \& Reisenegger, 1992) which may decay faster, thereby accelerating 
the decay of
the large scale dipolar field too. 
Since we present here only results of one--dimensional calculations 
we cannot modelize the process of the generation of small--scale field modes 
and their
interaction with the dipolar mode. 
Instead we mimic the effect of the Hall drift upon the field decay of its 
dipolar mode
by using an effective electrical conductivity 
\vspace{-0.2cm}
\begin{equation}
\sigma_{\rm eff}=\sigma_{\rm ohm}/\sqrt{1+(\omega_B \tau)^2}, 
\vspace{-0.1cm}
\end{equation}
where $\omega_{B}$ is the gyrofrequency of electrons and $\tau$ is their 
relaxation time.
This simple expression for $\sigma_{\rm eff}$ reproduces the two limiting 
cases of a weak
field (i.e., $\sigma = \sigma_{\rm ohm}$) and of a  strong field 
(i.e.,  $\sigma = \sigma_{\rm ohm}/\omega_{B} \tau$). 
The $\sigma_{\rm ohm}$ takes into account the different collision mechanisms 
of the
electrons that can appear in the crust (see Urpin et al., 1998). 
The metric function $\Phi(r)$ is related by Einstein's equations to the 
density and
pressure distribution within the crust and describes the general relativistic 
effects
which decelerate the field decay (Geppert, Page \& Zannias, 1999). 
We solve the induction equation simultaneously with the equations of the 
thermal evolution
by use of a Crank--Nicholson scheme.
The spin-history is calculated adopting the simple dipolar radiation losses 
equation.

\vspace{-0.2cm}
\section{Results}

\begin{figure}
\plotfiddle{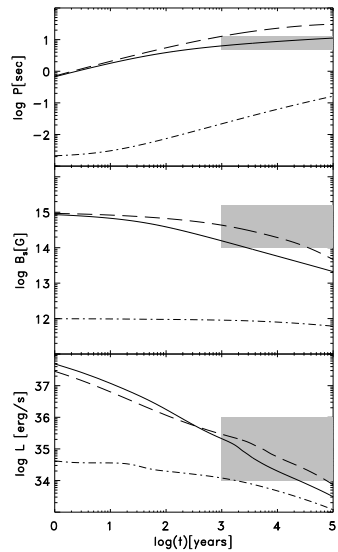}{6cm}{0}{50}{50}{-100pt}{-30pt}
\caption{Evolution of $P$, $B_s$, and $L$ for a model with a stiff EOS, 
standard cooling scenario and an initial field of $B_{s0} = 10^{15}$G 
which penetrates initially the crust up to densities $\rho_0 = 10^{13}$
g cm$^{-3}$ (full lines) and $10^{14}$g cm$^{-3}$ (dashed lines). 
For comparison we show by the dot dashed lines the evolution for the same 
model $B_{s0} = 10^{12}$G and $\rho_0 = 10^{14}$g cm$^{-3}$.
The shaded areas indicate the regions of typical values observed for magnetars.}
\end{figure}

\begin{figure}
\plotfiddle{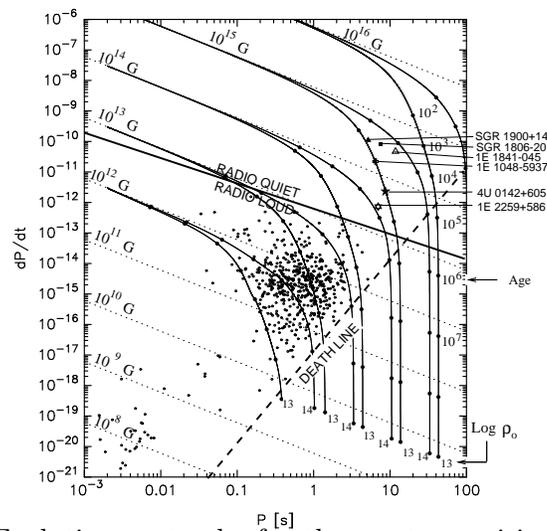}{6cm}{0}{40}{40}{-130pt}{-75pt}
\caption{Evolutionary tracks for the most promising model: stiff EOS and 
standard cooling. The diagram contains the positions of all pulsars (dots)
with $\dot P$ measurements from the Princeton Pulsar Catalog, the `pulsar 
dead line' as well as `radio quiescence line' above which the generation 
of the magnetospheric pair plasma is 
quenched by photon splitting. The positions of SGRs and AXPs are shown.
The ages of the models are indicated by the fat dots. $log(\rho_0)$ 
is indicated on each curve. Notice that a larger $\rho_0$ implies slower 
decay and thus larger periods are reached.}
\end{figure}

The evolution of the rotational period $P$ and its derivative $\dot P$, of the 
surface MF $B_s$ and of the luminosity $L$ for the  model which most 
promisingly coincides
with observations is represented in figures 1 and 2. 
The evolution of magnetars differs qualitatively from that of standard NSs. 
While the latter ones reach isothermality after a few $100$ yrs, the decay of 
an
ultra--strong MF in the crust causes an enormous heat release which can be 
appreciated
comparing the evolution of the luminosity of a standard NS with that of 
magnetars
(lower panel of figure~1). 
The delicate interplay between Joule heating, heat transport and neutrino 
cooling
results in significant temperature gradients in the whole crust for at least 
10$^4$ years.
To describe the observations the field decay must be, on the one hand, slow 
enough to
lead to large rotational periods and keep magnetic energy stored in the crust 
for a long
time but, on the other hand, fast enough to produce an intense Joule heating 
and to cluster
the final rotational periods around 8 s (see Colpi, Geppert \& Page 1999). Those delicate requirements can be 
accomplished
by a crustal MF. 

\vspace{-0.2cm}
\section{Discussion}

The assumption of a crustal field is not in contradiction with observations of 
AXPs,
provided the EOS is stiff enough and a standard cooling scenario applies. 
Furthermore, the initial MF strength should lie in a rather narrow range 
around $10^{15}$G.
This result may inspire speculations about a possible bimodal distribution of 
initial MFs
of NSs. 
The majority of NSs is born with canonical fields of $10^{12 \pm 1}$ G while, 
probably due
to another field generation mechanism, the magnetars are born with $10^{15 \pm 
0.5}$ G.
The crustal field hypothesis predicts naturally the existence of a critical saturation field of a few $10^{15}$G. Above this threshold the magnetic energy is preferentially conveyed into mechanical energy by 
fracturing the crystallized crust (Thompson \& Duncan 1996).
The rather stable spin--down of AXPs indicates that their field is just below this saturation value.\\
Concluding we can state that the crustal field hypothesis, although simply describing the Hall drift and its consequences, provides a framework for explaining the thermal evolution, luminosity and period clustering of magnetars.

\end{document}